# Integrated systems approach identifies pathways from the genome to triglycerides through a metabolomic causal network

Azam Yazdani [1*], Akram Yazdani [2], Philip L. Lorenzi [3], Ahmad Samiei [4]

[1] Climax Data Pattern, Houston, Texas, 77030, USA

[2] Department of Genetics and Genomic Sciences, Icahn School of Medicine at Mount Sinai, New York, 10029, USA

[3] The Proteomics and Metabolomics Core Facility, Department of Bioinformatics and Computational Biology, The University of Texas MD Anderson Cancer Center, Houston 77054, Texas, USA

[4] Hasso Plattner Institute, 14482 Potsdam, Germany

Address correspondence to:

Azam Yazdani, PhD

Climax Data Pattern, Houston, Texas, 77030, USA

a.mandana.yazdani@gmail.com

**Abstract**

Introduction: To leverage functionality and clinical relevance into understanding systems biology, one needs to understand the pathway of the genetic effects on risk factors/disease through intermediate molecular levels, such as metabolomics. Systems approaches integrate multi-omic information to find pathways to disease endpoints and make optimal inference decisions.

Method: Here, we introduce a multi-stage approach to integrate causal networks in observational studies and GWAS to facilitate mechanistic understanding through identification of pathways from the genome to risk factors/disease via metabolomics. The pathways in causal networks reveal the underlying relationships behind observations, which do not play a significant role in more traditional correlative analyses, where one variable at a time is considered.

Results: We identified a causal network over the metabolomic level using the genome directed acyclic graph (G-DAG), to systematically assess whether variations in the genome lead to variations in triglyceride levels as a risk factor of cardiovascular disease. We found *LRRC46* and *LRRC69* harboring loss-of-function mutations have significant effect on two metabolites with direct effects on triglyceride levels. We also found pathways of *FAM198B* and *C6orf25* to triglycerides through indirect paths from metabolites.

Conclusion: Integrating causal networks with GWAS facilitates mechanistic understanding in comparison to one-variable-at-a-time approaches due to accounting for relationships among components at intermediate molecular levels. This approach is complementary to experimental studies to identify efficacious targets in the age of big data sets.

## Introduction

Genome-wide association studies (GWAS) and recently whole genome sequence (WGS) studies have been widely conducted in humans with the goal of identifying genetic factors predictive of disease. Despite the extensive discovery of those studies, much of the genetic contributions to complex phenotypes remain unexplained. Furthermore, conclusions of noticeable numbers of genetic studies have not been in agreement with clinical presentation in an individual patient [1]. A key attribute for increasing confidence in potential clinical validity of gene variation with risk factors and disease end-points will be the development of assays with more direct mechanistic link. It is biologically meaningful that with a chronic systemic disease, molecular signals more proximal to the disease process may serve as strong biomarkers [2] and as a result, identify more stable pathways from the genome to disease risk factors and end-points. Therefore, integration of information in orthogonal data from different omics provides mechanistic understanding and has attracted attentions [3].

The metabolome is the end product of gene-environment interactions, Figure 1, and may be risk factors for future disease or biomarkers of current disease processes [4] [5] [6]. Metabolites can serve as intermediate phenotypes for genomic studies to illuminate mechanisms underlying of a specific SNP/gene, identify biological pathways linking the genome to disease, and discover valuable clinical biomarkers [2]. Therefore, integration of genetics and metabolomics holds potential for elucidating mechanisms for deciphering chronic disease.

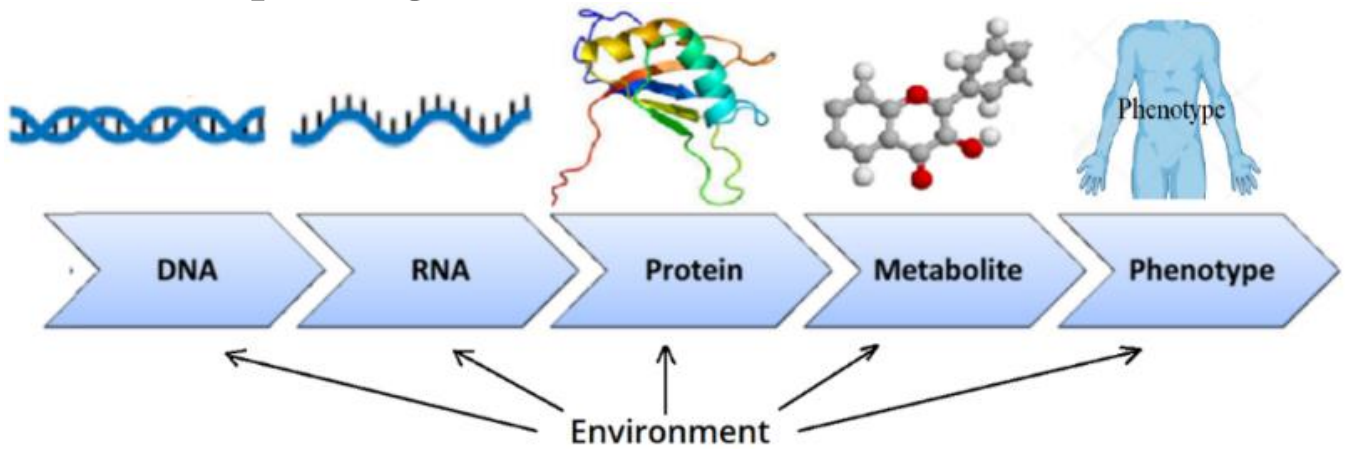

**Figure 1**. Metabolomics an integrated readout of many biological processes, such as genetic, transcriptomic, and proteomic variation, can characterize and recognize metabolic signatures of common chronic diseases.

Integration of data at different omics is challenging due to having largescale datasets and relationship between components. Most of the attempts at large scales are based on one component at a time that do not take into account the underlying relationships among components. Identification of causal networks based on Mendelian randomization and Bayesian graphical modeling is an established approach to discover relations among components of interest and reduce the risk of false positive discovery [8 9]. In this study, we introduce an integrative approach to identify pathways from the genome to risk factors via metabolomics. To provide insights into underlying relationships among metabolites, a metabolomic causal network is identified by leveraging genome information from using the G-DAG algorithm[10] (genome directed acyclic graph). Using the metabolomic causal network, we determine how genome variation leads to variability in risk factor levels through metabolomics. This approach reduces the spurious identification in comparison to one-metabolite-at-a-time approaches due to revealing relation among metabolites and identifying confounders at metabolomic level. To the best of our knowledge so far, no one has systematically analyzed metabolomic data to identify pathways from the genome to risk factor/disease via metabolomic networks. Here, we introduce a system approach to this aim and present an application.

**Mendelian randomization/instrumental variable**: Genetic markers have been employed in multiple studies to prevent the analyses being confounded [8 11]. This is called Mendelian randomization or instrumental variable (IV) approach which seeks to find a randomized experiment embedded in an observational study (Yazdani et al., 2014). IV approaches are based on some assumptions, see for example [13]. To hold the assumptions, the conventional approach is to find a genetic variant strongly associated with the variable of interest which may not be pragmatic at largescale metabolomic studies, see for example [14]. To overcome this challenge, the G-DAG algorithm generates strong IVs through extracting information from multiple variants and allocates multiple independent IVs to each metabolite, to make overall IVs even stronger, Figure 2.

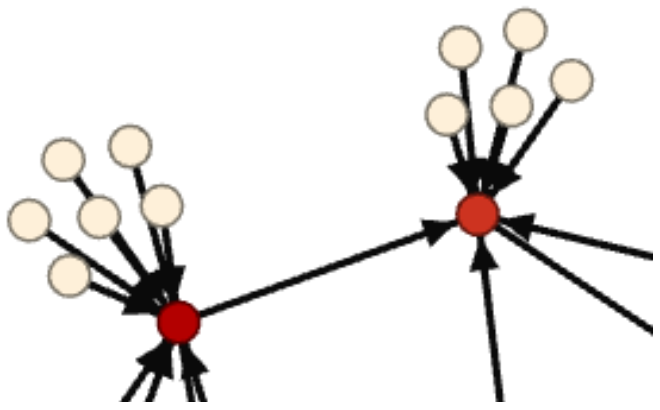

**Figure 2**. Visualization of some features of the G-DAG algorithm. White nodes indicate generated IVs from multiple SNPs. Brown nodes indicate metabolites. The aim is identification of underlying relationship between the metabolites (browns) using the IVs (whites).

The G-DAG algorithm combines this IV approach with Bayesian graphical modeling to identify a causal network over metabolites in observational studies. Using a metabolomic causal network, the approach below identifies pathways from the genome to risk factor/disease:

1. **Underlying relationships among metabolites:** Instead of analyzing an individual metabolite at a time, we identify a metabolomic causal network to infer underlying metabolomic relationships in an observational study. In the metabolomic network, nodes represent metabolites linked by directed edges which represent cause and effect relationships based on Mendelian principles. These relationship/pathway identification has multiple advantages, such as determining the role of each metabolite and its influence on the other metabolites as well as distinguishing intervention targets from disease predictors [15].

2. **Metabolites with direct effect on risk factors/disease:** Using the identified relationships among metabolites and application of structural equation modeling, we find metabolites with direct effect on risk factors (Yazdani et al., 2016a). Given the set of metabolites with direct effect on the risk factors, the rest of metabolites in the study do not have a significant effect on the risk factors. Therefore, a focus on the set of metabolites with direct effect on a risk factor is sufficient to know about the risk factor from the metabolomic level.

3. **Gene-metabolite relationship:** Through genome analysis with metabolites as phenotypes, we investigate gene-metabolite relationships.

After taking these steps, we can focus on genetic variations that exert their function through metabolomics which bridge gene effects to clinical end points. These pathways are identified after overcoming co-linearity at metabolomic level (step 1). Therefore, they facilitate mechanistic understanding and generate more efficacious hypotheses for clinical experiment.

**Case Study**

We focused on plasma triglycerides as a risk factor of cardiovascular disease. For the genome analysis, we focused on loss-of-function (LoF) mutations defined as sequence changes caused by single nucleotide variants or small insertions and deletions, which are predicted to result in a non-viable transcript or greatly truncated protein product [16].

**Study sample:** Genomic data and serum metabolites were available on a subset of the Atherosclerosis Risk in Communities (ARIC) study [17], 2,479 African-American (range in age from 45-64 years) who were randomly sampled from Jackson, Mississippi field center. Focusing on participants from the same race and region, we assume to overcome environmental confounders which may affect on metabolites, such as population-to-population and regional dietary variations. In addition to metabolites and dense genetic marker data, multiple risk factor phenotypes related to health and chronic diseases including plasma triglycerides were measured.

Metabolic profiling was completed in June 2010 carried out on fasting serum samples stored at -80 degrees centigrade since collection at baseline in 1986–1987. A total of 602 metabolites were detected and semiquantified by Metabolon Inc. (Durham, North Carolina) using an untargeted gas chromatography-mass spectrometry and liquid chromatography-mass spectrometry-based quantification protocol [18]. Metabolites were excluded on the basis of 3 criteria. First, more than 50% of the samples had missing values. Second, they had unknown chemical structures. Third, the metabolites or any transformation of them did not follow normal distribution. After this assessment, a total of 122 named metabolites were included in the study. Some information, such as the name and pathway of these metabolites are provided in Supplementary, Table S6.

Common single nucleotide polymorphisms (SNPs) were genotyped using the Affymetrix platform (version 6.0) consisting of 1,034,945 common variants spread across the genome. Variation across this data set was extracted and used to identify a metabolomic causal network.

Sequencing data of the protein-coding regions of the genome were also available; the annotations were captured by NimbleGen's VCRome2.1 (Roche NimbleGen), and the captured exons were sequenced using Illumina HiSeq 2000. The Burrows-Wheeler Aligner was used to align sequences to the hg19 reference genome [19]. Allele calling and variant call file construction were performed using the Atlas2 suite [20] (Atlas-SNP and Atlas-Indel). Variants were annotated using ANNOVAR [21] according to the reference genome GRCh37 and National Center for Biotechnology Information Reference Sequence.

**Identification of the metabolomic network:** We identified the metabolomic network over 122 metabolites using the G-DAG algorithm (Yazdani *et al.*, 2016b). The G-DAG algorithm first utilizes hierarchical clustering to measure linkage disequilibrium using square of correlation [22] and determine proxies from SNPs that are nearly perfectly correlated (>0.80) with others. Information from 1,034,945 SNPs scattered across the genome was extracted to generate multiple IVs. Over all of the chromosomes, 788 IVs were selected to identify the causal network. Minimizing the average structural Hamming distance (Tsamardinos et al., 2006), the tuning parameter was fit at level 0.001. In total 353 genome IVs remained in the model to identify the metabolomic network over 122 metabolites. The output is depicted in Figure 4 a. The employed IVs explained variation in the metabolites even up to 96%; although this result varied depending on the number of IVs influencing a specific metabolite. More details of the G-DAG algorithm are provided in Supplementary.

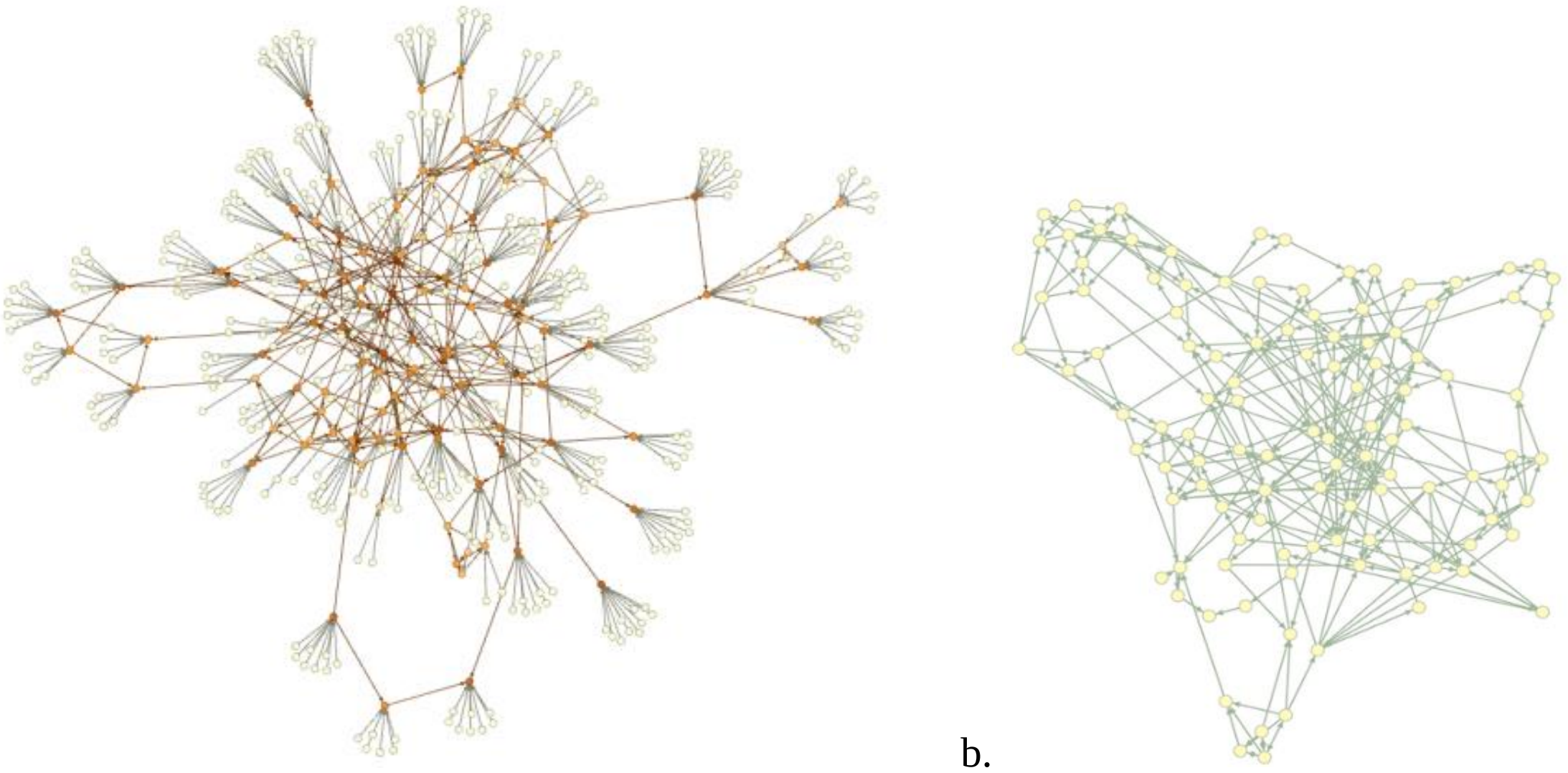

a. b.

**Figure 4. a**. Identified metabolomic causal network using the G-DAG algorithm established in Mendelian randomization and Bayesian network modeling. Pale nodes represent genome IVs which explained even up to 96% variation in metabolites. Orange nodes represent metabolites. **b**. The metabolite relationships from Figure 4a without depicting the IVs. Each link represents relationship between two corresponding metabolites when effects from other metabolites are excluded. Directions are identified based on Mendelian principles using genome IVs.

Note that the genome IVs were employed as a tool to identify the underlying metabolomic relationships. Therefore, after building the network, we focus on the metabolomic relationships, Figure 4b. In the metabolomic network, directions are identified robustly using Mendelian principles and represent cause and effect relationships. For further analysis of the metabolomic network, such as modules and individual properties of the metabolites, see [24].

**Metabolite-triglyceride pathways**: An extension of the G-DAG algorithm [25] was conducted to identify metabolites with direct effects on triglyceride levels and distinguish them from those with indirect effect, Figure 5. Using the underlying relationships between metabolites and triglycerides, we employed structural equation modeling and measured the causal effects. For more details see Supplementary, section 4. Nine metabolites were identified with direct effect on triglycerides at significance level 0.001; the estimated causal effects are presented in Supplementary, Table S2. The effect of the other metabolites to triglyceride levels is through the set of metabolites with direct effect on triglycerides. For the analysis, log transformation of triglyceride levels was adjusted for covariates including age and principle components for population stratification applying a linear regression. The analysis after including body mass index (BMI) in the set of covariates did not show significant effect of glutamate, glycine, and deoxycarnitine on triglyceride levels.

The systems approach applied facilitates distinguishing between direct and indirect metabolite-triglyceride pathways. Some of the novel findings of this analysis have already been validated clinically (Yazdani et al., 2018). While through an association study 21 metabolites out of the 122 are found associated with triglyceride levels, applying the systems approach identified that only 6 of them (after adjustment for BMI) have direct effect on triglyceride levels. Therefore, the systems approach applied here provides efficacious targets for intervention compared to association studies. For the results of the association study, see Supplementary, Table S3.

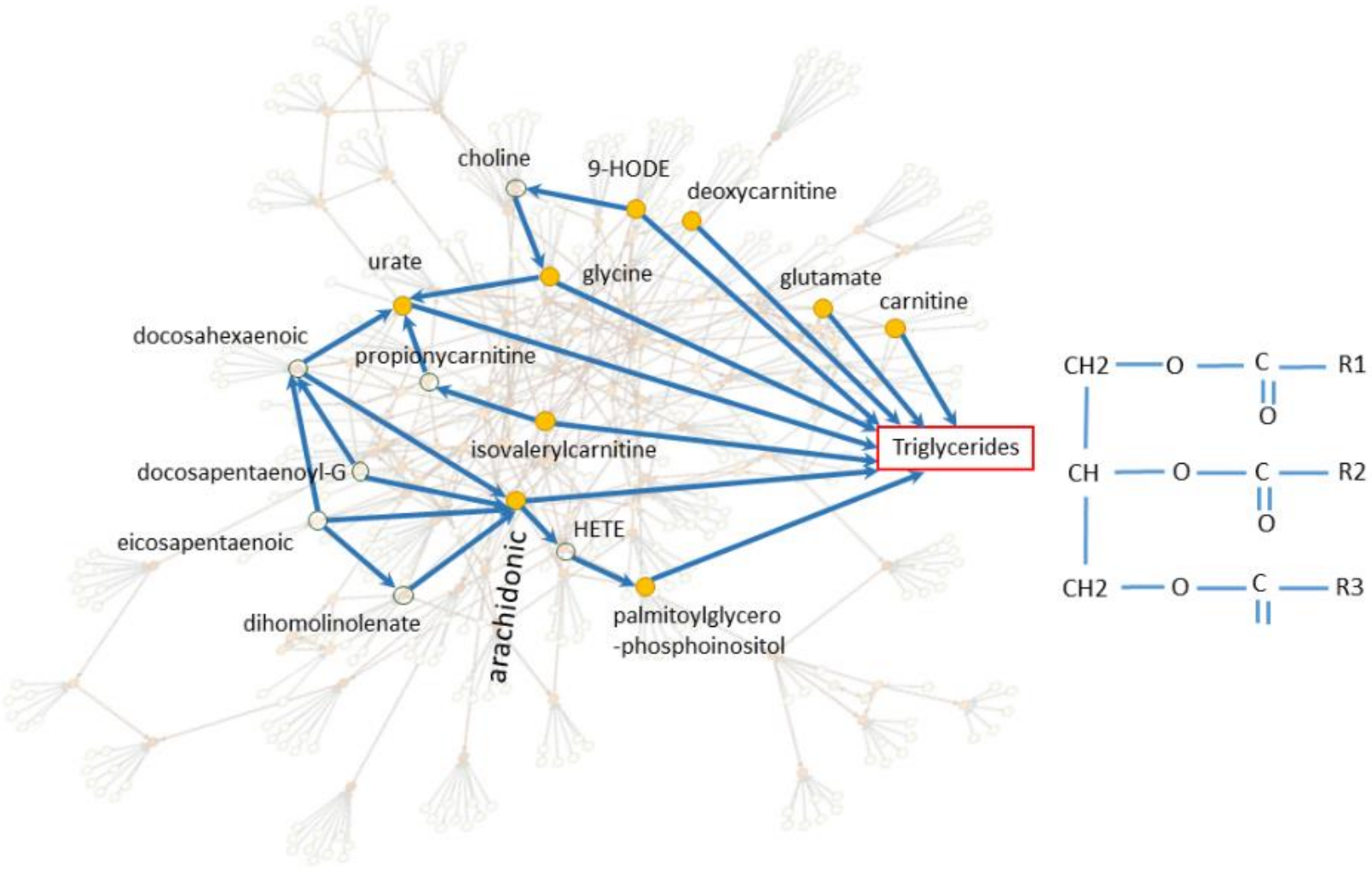

**Figure 5**. Relationship among metabolites with direct effect on triglyceride levels. Arachidonic acid has the largest effect on triglycerides. No feed-back loop (i.e. cycles) was identified between these metabolites and triglyceride levels. The relationships among metabolites are identified from the genomic-metabolomic network depicted in the background to emphasize that the directions are identifies based on Mendelian principles.

**LoF mutation-metabolite pathway:** LoF variants included in this study were defined as premature stop codons occurring in the exon, essential splice site disrupting, and indels predicted to disrupt the downstream reading frame. Seven genes with significant effect ($P_val < 1.3 \times 10{-}7$) on metabolites in our analysis are integrated with the metabolomic network, Figure 6 and Supplementary, Table S4. For the analysis, single-variant tests for the single variants and gene-based burden tests for the genes were utilized [26] to investigate the relationship with the individual metabolites.

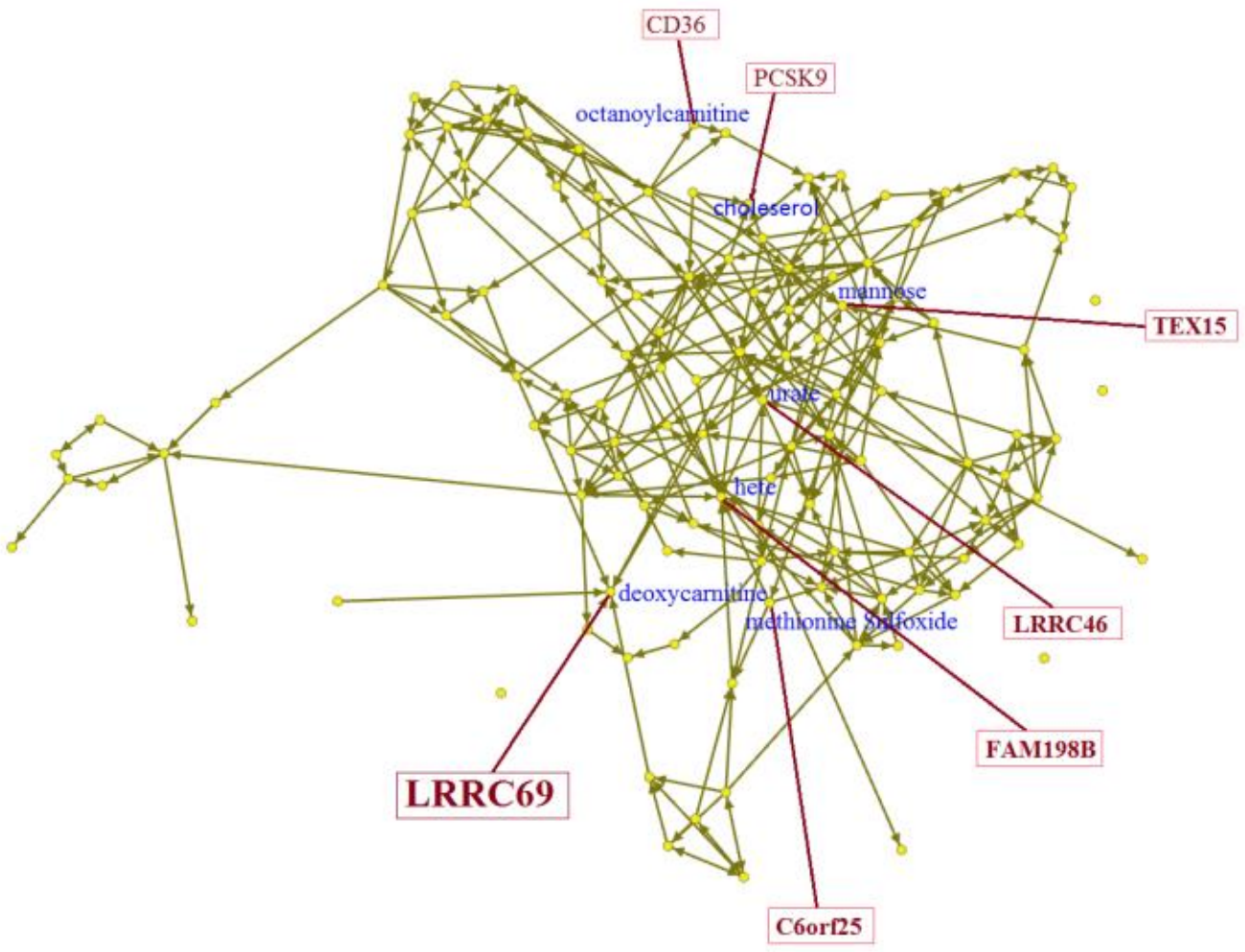

**Figure 6**. Gene-metabolomic network. The genes harboring LoF mutations are depicted on the metabolomic network in red rectangular. The strongest relationship is for *LRRC69*, which is depicted in a bigger size than the other genes.

## Identified genome-metabolite-triglyceride pathways

Through combining the results of the three steps above, we identify pathways from the genome to plasma triglycerides via metabolomic network. By inspecting the metabolites with direct effect on triglyceride levels and those influenced by LoF mutations, we identified two direct pathways linking the genome to plasma triglycerides: One path from *LRRC46* through metabolite urate (*LRRC46* → Urate → Triglycerides), and another path from *LRRC69* through metabolite deoxycarnitine (*LRRC69* → Deoxycarnitine → Triglycerides). These pathways are visualized in Figure 7.

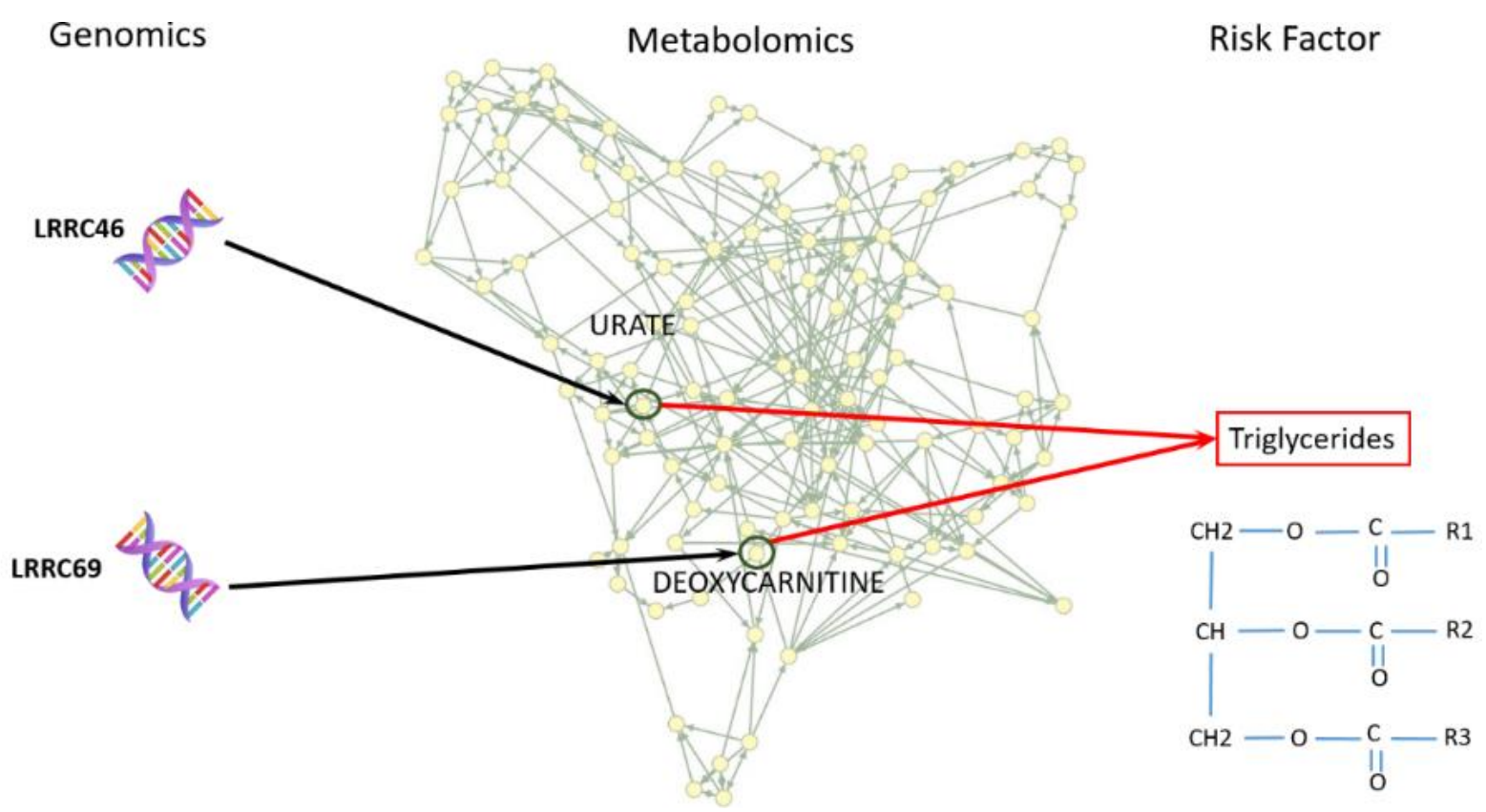

**Figure 7**. Two identified direct pathways from *LRRC46* and *LRRC96* to plasma triglycerides via metabolites urate and deoxycarnitine respectively.

In Figure 8 below, blue pathways depict metabolites with influence on urate and deoxycarnitine. In the left panel, propionylcarnitine and isovalerate are lipids involved in fatty acid metabolism; docosahexaenoate is an essential fatty acid; glycine is an amino acid involved in glycine, serine, and threonine metabolism; and finally gamma-glutamylthreonine is a peptide in gamma-glutamyl metabolism. In the right panel of Figure 8, 3-carboxy-4-methyl-5-propyl-2-furanpropanoate, azelate, and laurate are fatty acids; citrate is a component of the tricarboxylic acid cycle that is central to energy metabolism; and trans-4-hydroxyproline is a modified amino acid associated with the urea cycle and is thought to be associated with oxidative stress.

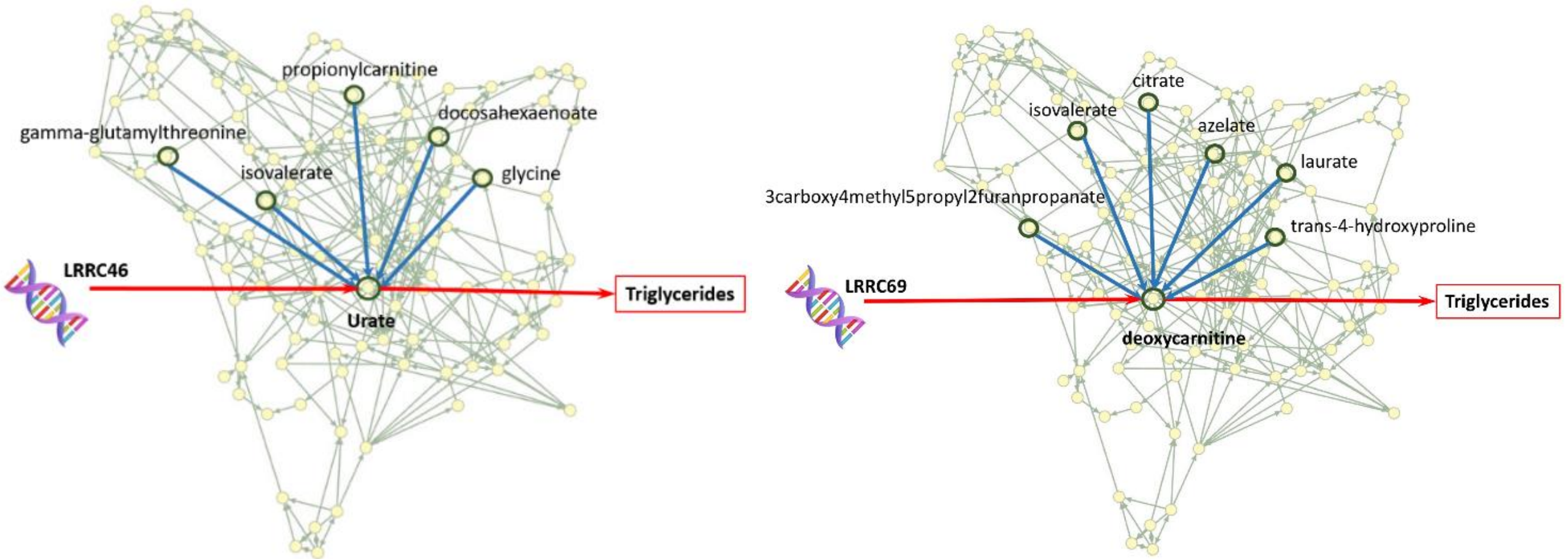

**Figure 8**. Red pathways: Direct pathways from the genome to triglycerides via urate and deoxycarnitine. Blue pathways: Metabolites that influence urate and deoxycarnitine at metabolomic network.

In addition to the above direct pathways from the genome to triglycerides via metabolomics, two indirect pathways are identified to triglycerides. Genes *FAM198B* and *C6orf25* influence metabolite HETE with indirect effect on triglyceride levels. These pathways are through metabolite palmitoylglycerophosphoinositol. In Figure 9, we see that the identified pathways to triglycerides is not through arachidonic acid. Rather, arachidonic acid influences this path through HETE. These results are identified using the metabolomic causal network.

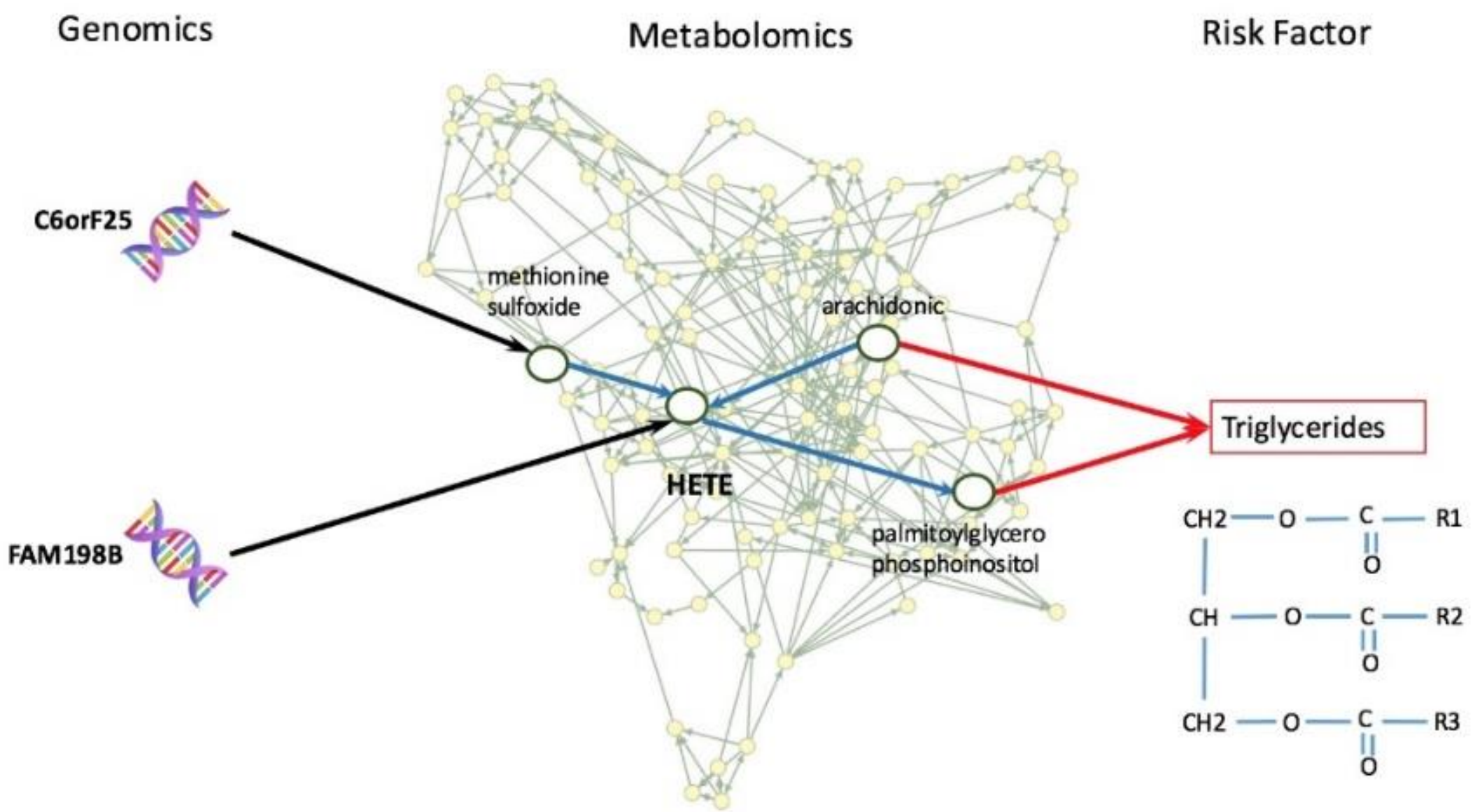

**Figure 9**. Indirect pathways from *FAM198B* and *C6orf25* to triglycerides through metabolite HETE and palmitoylglycerophosphoinositol.

The identified pathways through the systems approach introduced here may facilitate understanding underlying mechanism and generate hypotheses for further experimental studies. Through a standard GWAS analysis, we investigated genome-triglyceride relationship and no variant with significant effect on triglyceride levels was identified. The results are provided in Supplementary, Table S5.

## Discussion

The genetics of complex diseases, such as cardiovascular disease, produce alterations in the molecular interactions of intermediate phenotypes, such as metabolites, which contribute to early disease-related changes [27 28]. The collective effect in cellular pathways may become clear through integrated approaches to identify underlying mechanisms. Powerful and advanced analytic strategies are required to integrate largescale data in systems biology for the elucidation of pathways across human biological levels. We introduced a systems approach for pathway identification by integrating data at three different biological levels using causal networks. Causal networks compatible with structural equation modeling improve the power of discovery by reducing the influence of phenotype-phenotype associations to illustrate underlying relationships [7 29].

The introduced approach is toward mechanistic understanding through pathway identification linking the genome to health/disease. The pathways reveals the underlying relationships behind observations [303112], which do not play a significant role in more traditional correlative analyses. We first construct a causal network over metabolomics using IVs/Mendelian randomization [7]. Through a metabolomic causal network, we not only account for association between metabolites but also confounders at metabolomics. Second, we take an improvement in understanding the role of metabolites in quantitative risk factors, here triglycerides [7]. This step filters the number of metabolites to a subset that impact triglyceride levels. In addition, the visualization of underlying relationships reveals even more information.

Through integration of the metabolomic causal network and genome association study, two pathways were identified, from the genome—leucine rich repeat containing 46, LRRC46, and leucine rich repeat containing 69, *LRRC69*—linked to plasma triglycerides via the metabolites urate and deoxycarnitine, respectively. Those metabolite-triglyceride connections are consistent with known biochemical information. For example, urate is biochemically linked to the thioredoxin system, which mediates cellular redox balance and has been associated with triglyceride levels [32]. Similarly, deoxycarnitine lies at a biochemical crossroad between triglyceride metabolism and fatty acid oxidation, and carnitine metabolism has been implicated in the regulation of triglycerides [33]. Further, carnitine is essential for β-oxidation of long-chain fatty acids, and metabolic enzymes involved in carnitine biosynthesis mediate a decrease in fatty acid oxidation and increase in glycolysis in heart failure progression [34].

The genetic components *LRRC46* and *LRRC69* provide new mechanistic insights into the regulation of triglyceride levels. Experimental validation will need to be conducted to assess the contributions of *LRRC46* and LRRC69 to the modulation of triglyceride levels, but our new approach allows such validation experiments to be focused. For instance, instead of measuring only triglyceride levels as an endpoint in a model biological system, we can additionally measure levels of uric acid, deoxycarnitine, and other metabolites known to be associated with each of them, such as thioredoxins. For example, since thioredoxins have been found to mediate cardioprotection, it may be possible to implicate *LRRC46* in cardioprotection upstream of thioredoxins regulation.

The leucine-rich repeat (LRR) structural motif is characterized by the α/β horseshoe fold, composed of 20-30 hydrophobic amino acid stretches of leucine. LRRs mediate protein-ligand interactions and in the case of cascade interaction model in fatty-acid-uremic toxin-drug system, in which long-chain fatty acids concentrations are increased, cascade displacement of bound drugs occurs by a competitive inhibitor, such as CMPF and uremic toxins containing an indole ring [35]. Previous studies reveal a relationship between leucine-rich repeat (LRR) and cardiovascular disease and triglycerides [3637]. The pathways identified here strengthen those associations and provide a plausible mechanism for them.

The approach presented here provides a step toward addressing challenges in modern biomedical research, such as largescale data sets, highly correlated phenotypes, and integrating information at different biological levels. Additionally, future method development will include genome variation effects on multiple metabolites, hypothesized pleiotropy in GWAS/WGS, to provide further insights into the mechanistic underpinnings of chronic diseases.

**Acknowledgments**: This research is supported by a training fellowship from the Keck Center for Interdisciplinary Bioscience Training of the Gulf Coast Consortia (Grant No. RP140113).

**Author contributions**

AY (Azam Yazdani) carried out the analysis and wrote the manuscript.

AY (Akram Yazdani) significantly contributed in writing the manuscript and essential comments which improved the manuscript.

PL contributed in the interpretation of some of the results provided in the discussion.

AS carried out the visualizations.

**Competing interests**

The author(s) declare no competing interests.

**Data availability**

The data is available through dbGaP https://www.ncbi.nlm.nih.gov/gap/?term=aric and ARIC study upon the request.

**Integrated systems approach identifies pathways from the genome to triglycerides through a metabolomic causal network**

Azam Yazdani[1*], Akram Yazdani[2], Philip L. Lorenzi[3], Ahmad Samiei[4]

## Supplementary

**Content**

### 1. Causal Inference in Observational Study

In observational studies, where intervention is not possible, causal inference is derived by identifying the assignment mechanism (*AM*), i.e. causal relationships, underlying the observations (1)(2)(3)(4). The *AM* represents how different levels of a response variable are assigned and which covariates are involved. The illumination of the *AM* is carried out using knowledge about response variables ($K_R$) and is required for any causal inference. This assumption is formalized by $AM(K_R)$ and is called causal parameter (4). Any causal quantity is implicitly or explicitly conditioned on the causal parameter $AM(K_R)$. The causal parameter represents the necessity of illumination of the *AM* by $K_R$.

In this study, the *AM* is illuminated by the fact that inherited genomic variation is a causal factor for variations in the metabolomics level (an upper granularity) and not the other way around. This is the applied knowledge about metabolite response variables ($K_R$) to illuminate the *AM*. The *AM* or underlying relationship is estimated using different approaches, such as propensity score, Matching, or causal networks. Here, we utilize causal networks and illuminate the *AM* by a directed acyclic graph (3). We then graphically identify confounders among covariates to measure causal effect sizes. For the definition of confounders and covariates see (5).

This section is published previously as supplementary of (6). It is provided here for the readers' convenience.

## 2. The G-DAG algorithm

**First**, there is a reduction in the number of SNPs using hierarchical clustering and the square of the correlation measure of linkage disequilibrium, e.g. see (7). **Second**, the genomic information was summarized using the set of principal components applied to each chromosome to create IVs. **Third**, the set of IVs responsible for >80% of the variation over each chromosome was selected. **Fourth**, the G-DAG algorithm determines genome-wide IVs that are significantly related to each metabolite excluding the effect of other metabolites and IVs in the model. **Fifth**, The G-DAG algorithm dentifies a topology over metabolomics, an undirected metabolomics Bayesian network determined through metabolite analysis. Sixth, using the IVs and Bayesian rules, the G-DAG algorithm determines directions over the metabolomic topology. The analysis can be carried out either in a constraint-based or score-based framework. We have implemented the G-DAG algorithm in a constraint-based framework (8). Note that in the G-DAG algorithm, using the same data twice is avoided since unsupervised statistical methods are employed to extract information from the genotype data.

The causal network is a representation of following properties (Metabo paper):

$$M_i \perp M_r \mid (S_{M_i}, AM(K_R)), \text{ for } M_r \notin S_{M_i}, \qquad (1)$$

where $M_j, j = 1,...,m$ stands for a metabolite in the system and *m* is the number of metabolites. $S_{M_i}$ includes metabolites with an arrow in node $M_i$ in the network and $AM(K_R)$ stands for causal parameter (4). Representation of underlying relationships is a representation of Assignment Mechanism (*AM*) which can be illuminated using Knowledge about Response ($K_R$). Conditioning the properties (1) on the causal parameter $AM(K_R)$ explicitly represent that underlying relationships are taken into account to analyze data in a causal setting.

**3. Features of the G-DAG algorithm**

The G-DAG algorithm utilizes principal component analysis to extract near-complete information from across the genome and create strong and independent IVs. Principal components are powerful in capturing the effect of multiple loci (9).

Each of the IVs includes information from several SNPs/genes across the genome, therefore, they are much stronger than a single gene. Since information from each SNP is extracted, the G-DAG algorithm is able to create several hundreds strong instrumental variables and identify directions robustly over a metabolomics topology in large scale. The G-DAG algorithm takes the following steps to hold the underlying assumptions of Mendelian randomization/instrumental variable approach through following features:

1) Creating and employing strong instrumental variables through extracting information from multiple Variants across the genome;
2) Creating independent instrumental variables toward holding the assumption of validity;
3) Multiple independent instrumental variables for each metabolite, to make overall IVs even stronger.

Employed instrumental variables explain a majority of variations in each metabolite, even up to 96%. Since a majority of variations in each metabolite is explained by instrumental variables, the identified directions are robust (15). Utilizing independent IVs is toward overcoming pleiotropic effect of SNPs, LD structure, and even gene-gene interactions.

**4. Causal effect measurement**

We measured the effect of each metabolite on baseline triglyceride levels given the overall metabolomic-triglecerides network by

$$RF \mid AM(K_R) = \sum_{i \in S_{RF}} \beta_i M_i + U$$

where $M_i$ stands for *i*th metabolite and $S_{RF}$ stands for a set of metabolites with direct effect on *RF*. The *RF* corresponds the risk factor triglyceride levels. *U* stands for all factors with effect on RF but independent from $M_i$ and $AM(K_R)$ stands for the causal parameter and conditioning on the causal parameter is corresponding to illuminating the assignment mechanism (AM)/causal relationships behind observations by gathering knowledge about response, here RF. By illumination of the *AM* through causal ntworks, we can identify confounders at the metabolomic level and consider them in the model to assume $U \perp M_i$, see graphical criteria in (3). As a result, we can estimate the effect of a specific metabolite on triglyceride levels, and the coefficients $\beta_i$ have causal interpretations. Note that equations written based on causal networks are called structural equation models. In other word, causal networks and structural equation modelings are compatible (3). For more details of causal effect measurement see (10) (11). A version of this section is published as supplementary in (6).

## 5. Materials and data measurments

**Study Population:** The Atherosclerosis Risk in Communities (ARIC) study is a prospective epidemiological study designed to investigate the etiology and predictors of cardiovascular disease (CVD). The ARIC study enrolled 15,792 individuals aged 45-64 years from four U.S. communities (Forsyth County, NC; Jackson, MS; suburbs of Minneapolis, MN; and Washington County, MD) in 1987-89 (baseline) and followed for four completed visits in 1990-92, 1993-95, 1996-98 and 2011-13. A detailed description of the ARIC study design and methods is published elsewhere (12). Basic cardiovascular disease risk factors were measured at each visit, and cardiovascular disease endpoints, including heart failure, were ascertained annually using telephone interviews and hospital medical record review. In ARIC, incident heart failure (HF) was defined as the first hospitalization or death from HF for those without a prior HF hospitalization. The diagnosis of HF was based on International Classification of Diseases, Ninth Revision (ICD-9) code 428.3. Individuals were followed up for events from baseline through December 31, 2011; and those who were lost to follow-up were censored at the date of last contact. Incident type 2 diabetes was defined in diabetes free participants at baseline and having a follow-up visit until visit 4 with either: (i) a fasting glucose ≥7.0 mmol/L, (ii) a non-fasting glucose ≥11.1 mmol/L, (iii) use of a diabetes medication, or (iv) self-reported physician diagnosis.

**Metabolome Measurements:** Metabolite profiling was measured using fasting serum samples collected at the baseline examination in 1987-1989 among ARIC European- and African-Americans. For the discovery African-American samples, a total of 602 metabolites were detected and quantified by Metabolon Inc. (Durham, USA) using an untargeted, gas chromatography-mass spectrometry and liquid chromatography-mass spectrometry (GC-MS and LC-MS)-based metabolomic quantification protocol (*2, 3*). Detection of metabolites were completed in June 2010. Metabolites were excluded if: i) Missing values more than 50% of the samples, ii) Unknown chemical structures, and iii) Not following normal distribution. After these assessments, a total of 122 metabolites were included in the present study. Replication samples were utilized to identify an optimal approach for imputation which was KNN. Therefore, missing values were imputed by KNN algorithm. Metabolite levels were transformed to normal with different transformation (not necessary log-transformation) prior to the analyses.

**Whole Exome Sequencing:** The annotated exome was captured by Nimblegen's VCRome2.1 (Roche NimbleGen, Madison, WI, USA) and the captured exons were sequenced using Illumina HiSeq 2000 (Illumina, San Diego, CA, USA). The Burrows–Wheeler Aligner was used to align sequence to the hg19 reference genome (13). Allele calling and variant call file (VCF) construction was performed using the Atlas2 suite (14) (Atlas-SNP and Atlas-Indel).

Single nucleotide variants were excluded if they had a posterior probability less than 0.95, total depth of coverage less than 6x, allelic fraction < 0.1, 99% of reads in a single direction and homozygous reference alleles with < 6x coverage. Low-quality indels were excluded if they had minimum total depth < 30, allelic fraction < 0.2 for heterozygous variants and < 0.8 for homozygous variants and variant reads < 10x. Following statistical analysis, all significantly associated variants were validated using an orthogonal laboratory technology (i.e. Array-based genotyping, Sequenome genotyping, or Sanger sequencing). All reported results were validated with 100% accuracy.

**Figure A1.** Distribution of the number of LoF variants per individual in ARIC African-American population.

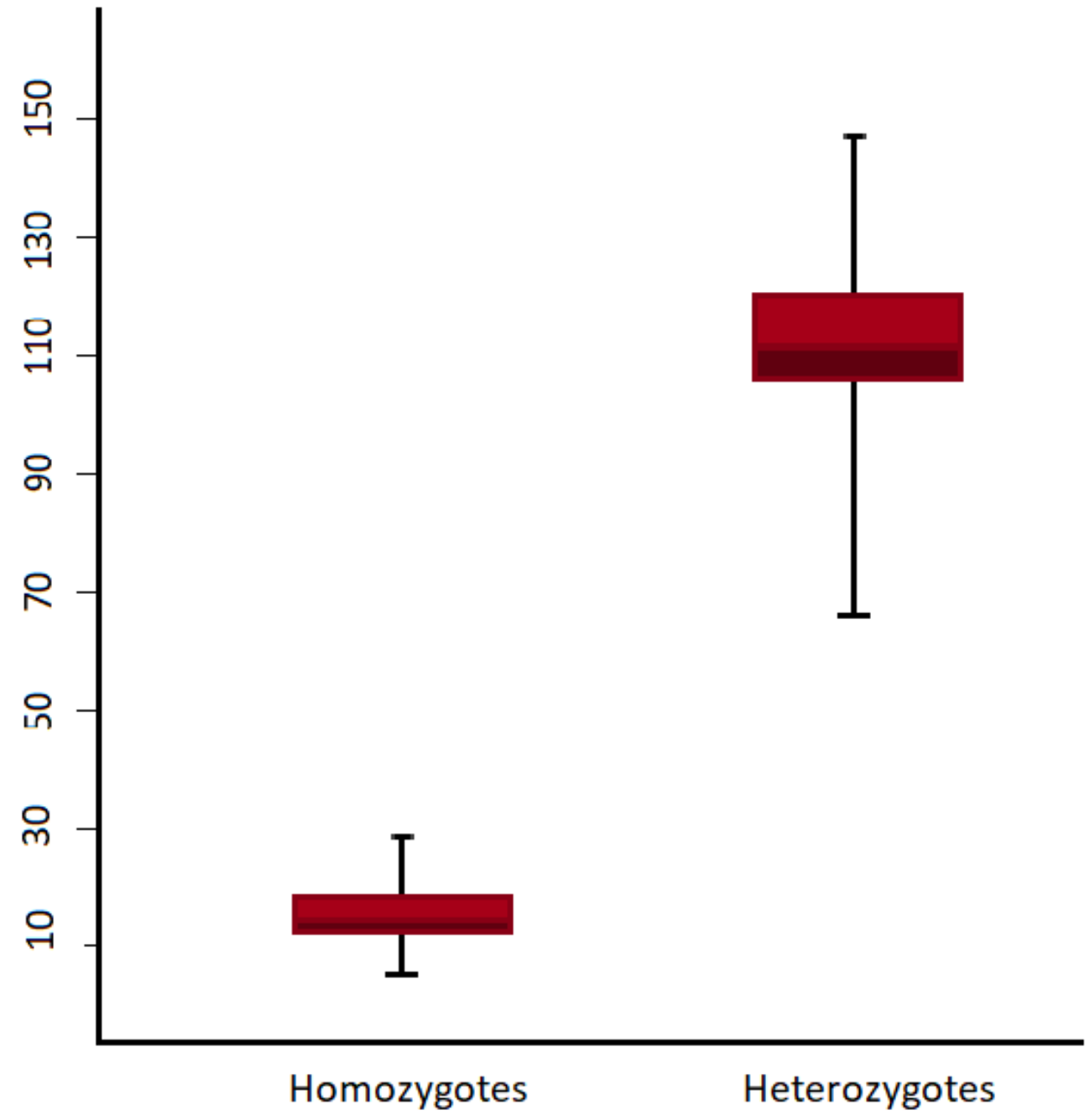

Mean values ± standard deviations for continuous variables.

**Table S1**. Baseline Characteristics of African Americans in ARIC with metabolomic and exome sequence data.

| **Characteristic** | **N = 1,361** |
|---|---|
| Age, years | 52.5 ± 5.6 |
| Male, n (%) | 464 (34.1) |
| Body mass index, kg/m2 | 29.8 ± 6.2 |
| Hypertension, n (%) | 705 (51.8) |
| Diabetes, n (%) | 190 (14.0) |
| Prevalent coronary heart disease, n (%) | 49 (3.6) |
| Current smoker, n (%) | 377 (27.7) |
| Systolic blood pressure, mm Hg | 126.9 ± 19.0 |
| Diastolic blood pressure, mm Hg | 80.2 ± 11.4 |
| HDL cholesterol, mg/dL | 56.1 ± 17.4 |
| LDL cholesterol, mg/dL | 135.2 ± 38.0 |
| Triglycerides, mg/dL | 106.2 ± 55.9 |
| Total cholesterol, mg/dL | 212.5 ± 40.0 |

**Table S2**. The causal network parameters and the effect sizes of nine metabolites with direct effect on plasma triglycerides.

| Metabolite | Out-degree | In-degree | Strength | Pathway | P-value | Effect Size[a] (SE) |
|---|---|---|---|---|---|---|
| Arachidonic | 1 | 4 | 50 | Lipid | 2.3e-17 | 0.17 (0.03) |
| Carnitine | 1 | 1 | 7 | Lipid | 1.4e-11 | 0.15 (0.04) |
| 9-HODE | 1 | 1 | 50 | Lipid | 1.4e-7 | 0.12 (0.03) |
| Palmitoylglycero-phosphoinositol | 1 | 5 | 8 | Lipid | 1.6e-6 | 0.10 (0.01) |
| Urate | 0 | 5 | 11 | Nucleotide | 2.2e-5 | 0.09 (0.01) |
| Isovalerylcarnitine | 2 | 0 | 20 | Amino acid | 2.0e-4 | 0.09 (0.02) |
| Glycine [b] | 4 | 2 | 20 | Amino acid | 4e-3 | -0.09 (0.02) |
| Deoxycarnitine [b] | 0 | 6 | 11 | Lipid | 1e-3 | -0.08 (0.03) |
| Glutamate [b] | 0 | 0 | 0 | Amino acid | 1e-3 | 0.07 (0.02) |

a. Effect sizes measured in standard deviation units to facilitate comparison.
b. Metabolite with no significant effect at level 0.0001 after adjustment for BMI

The analysis after including body mass index (BMI) in the set of covariates did not show significant effect of glutamate, glycine, and deoxycarnitine on triglyceride levels, which are noted with superscript “b” in Table S2.

More information can be obtained from further analysis of metabolomic network called causal parameters(15). In Table S2, in addition to p-value and effect size, three causal network parameters Out-degree, In-degree, and Strength are presented to investigate potential roles of the metabolites.

- In-degree represents the number of metabolites that influence a particular metabolite. At the metabolomic causal network the In-degree parameter has a range from 0 to 9. Metabolites with higher In-degree capture features of higher number of metabolites in the metabolomic network.

- Out-degree represents the number of metabolites influenced by a particular metabolite. At the metabolomic causal network the Out-degree parameter has a range from 0 to 8 at the metabolomic causal network. Metabolites with higher Out-degree influence higher number of metabolite in the network.

- Strength represents how strong a metabolite is connected to the network, ranges from 0 to 50 at the metabolomic causal network. The Strength zero (the lowest strength) means the metabolite is not connected to the network.

In comparison to the range of Out-degree in the metabolomic network (0 to 8), the study reveals that the metabolites with direct effect on triglycerides have very low Out-degree, which is corresponding to their low influence in metabolomic network.

There are other approaches for pattern recognition that can be brought in the biomedical research such a (16) (17)(18)(19)(20). These approaches can improve pattern identification among highly correlated traits at each omic.

**Table S3.** Assocciated metabolites with triglycerides at significance level 4e-4 (Bonferroni correction of 122 test for 122 metabolites) which carried out via linear regression.

| ID | **Metabolite** | **p-value** |
|---|---|---|
| 1 | arachidonate | 4.29E-26 |
| 2 | dihomolinolenate | 1.66E-12 |
| 3 | carnitine | 1.30E-11 |
| 4 | urate | 4.86E-10 |
| 5 | docosahexaenoate | 6.72E-10 |
| 6 | isoleucine | 7.65E-10 |
| 7 | palmitoylglycerophosphoinositol | 4.83E-09 |
| 8 | leucine | 1.11E-08 |
| 9 | adrenate | 5.32E-08 |
| 10 | 9HODE | 5.94E-08 |
| 11 | eicosapentaenoate | 1.25E-07 |
| 12 | isovalerate | 7.15E-07 |
| 13 | valine | 1.14E-06 |
| 14 | arachidonoylglycerophosphoinositol | 2.29E-06 |
| 15 | isovalerylcarnitine | 1.16E-05 |
| 16 | phenylalanine | 8.79E-05 |
| 17 | glycerol | 9.81E-05 |
| 18 | propionylcarnitine | 0.000181773 |
| 19 | kynurenine | 0.000247776 |
| 20 | glycine | 0.000257107 |
| 21 | glutamate | 0.0004107 |

**Table S4**. Genes harboring LoF mutation with significant effect on individual metabolites under study. The causal network parameters are measured from the metabolomic causal network. Variant includes chromosome:position:reference allele:alternative allele.

| Gene and Variant | Metabolite | Out-degree | In-degree | Strength | Pathway | P-value |
|---|---|---|---|---|---|---|
| LRRC69<br>8:92213022:T:A & 8:92212839:A:G | Deoxycarnitine | 0 | 6 | 11 | Lipid | 9e-16 |
| LRRC46<br>17:45913719:TG:T & 17.45911791:GT:G | Urate | 0 | 5 | 11 | Nucleotide | 1e-7 |
| FAM198B<br>4:159091864:G:A & 4:159091422:T:A | HETE | 2 | 5 | 17 | Lipid | 5e-9 |
| CD36<br>7:80300449:T:G | Octanoylcarnitine | 1 | 1 | 50 | Lipid | 4e-8 |
| PCSK9<br>1:55529215:C:A & 1:55512222:C:G<br>1:55524293:TG:T | Cholesterol | 0 | 3 | 29 | Lipid | 5e-9 |
| TEX15<br>8:30700833:TTC:T & 8:30694729:C:CA<br>8:30701535:C:A & 8:30705099:CTCTA:C<br>8:30702839:CTTTCA:C | Mannose | 1 | 1 | 14 | Carbohydrate | 9e-9 |
| C6orf25<br>6:31692558:C:T | Methionine sulfoxide | 3 | 1 | 13 | Amino acid | 3e-8 |

LoF variants included in this study were defined as premature stop codons occurring in the exon, essential splice site disrupting, and indels predicted to disrupt the downstream reading frame.

Potential roles of these metabolites in the metabolomic system are investigated through causal network parameters. The metabolites influenced by LoF mutations tabulated in Table 2 do not have important roles at the metabolomic network due to low number of Out-degrees and relatively higher number of In-degree. Later, we see both metabolites urate and deoxycarnitine are in direct paths to triglycerides, Figure 7. Interestingly, both of them have a high In-degree and zero Out-degree. We conclude the metabolites influenced by LoF mutations are mostly influenced by metabolomic system rather than influence the system. This may be interpreted as below: metabolites influences by LoF mutations cannot be so critical at metabolomics. Otherwise they will lead to debilitating disease or be inconsistent with life. This point can be considered and discussed for functional understanding although it may require further assessments.

**Table S5**: Six top lowest p-value from genome-triglyceride analysis. Triglycerides is adjusted for BMI. No variant with a significant effect on triglyceride levels is identified.

| SNP | Estimate | Std. Error | t-value | p-value |
|---|---|---|---|---|
| X9.21481483.G.A | 0.184157 | 0.061768 | 2.981432 | 0.002903 |
| X13.111367876.G.C | 0.195495 | 0.067422 | 2.899555 | 0.003777 |
| X17.67149973.G.A | 0.168886 | 0.060432 | 2.794665 | 0.005244 |
| X9.73152037.C.T | 0.925106 | 0.376876 | 2.454672 | 0.014184 |
| X16.90128376.G.A | 0.252941 | 0.116398 | 2.173062 | 0.02989 |

| X3.148559694.C.T | 0.539245 | 0.27711 | 1.945963 | 0.051796 |
|---|---|---|---|---|

**Table S6**. The super-pathway, sub-pathway, and measurement platform for 122 metabolites measured in African-Americans population in ARIC study.

| BIOCHEMICAL | SUPER_ PATHWAY | SUB_ PATHWAY | PLATFORM |
|---|---|---|---|
| **Glutarate** | Amino acid | Lysine metabolism | GC/MS |
| **p-cresol sulfate** | Amino acid | Phenylalanine & tyrosine metabolism | LC/MS neg |
| **Leucine** | Amino acid | Valine, leucine and isoleucine metabolism | LC/MS pos |
| **Isoleucine** | Amino acid | Valine, leucine and isoleucine metabolism | LC/MS pos |
| **Valine** | Amino acid | Valine, leucine and isoleucine metabolism | LC/MS pos |
| **gamma-glutamylalanine** | Peptide | gamma-glutamyl | LC/MS pos |
| **gamma-glutamylglutamate** | Peptide | gamma-glutamyl | LC/MS pos |
| **gamma-glutamylisoleucine** | Peptide | gamma-glutamyl | LC/MS pos |
| **gamma-glutamylleucine** | Peptide | gamma-glutamyl | LC/MS pos |
| **gamma-glutamylphenylalanine** | Peptide | gamma-glutamyl | LC/MS pos |
| **gamma-glutamylthreonine** | Peptide | gamma-glutamyl | LC/MS pos |
| **gamma-glutamylvaline** | Peptide | gamma-glutamyl | LC/MS pos |
| **Glycylleucine** | Peptide | Dipeptide | LC/MS pos |
| **Leucylalanine** | Peptide | Dipeptide | LC/MS pos |
| **Serylleucine** | Peptide | Dipeptide | |
| **Caproate** | Lipid | Medium chain fatty acid | LC/MS neg |
| **Myristate** | Lipid | Long chain fatty acid | LC/MS neg |
| **Myristoleate** | Lipid | Long chain fatty acid | LC/MS neg |
| **Palmitate** | Lipid | Long chain fatty acid | LC/MS neg |
| **Palmitoleate** | Lipid | Long chain fatty acid | LC/MS neg |
| **Margarate** | Lipid | Long chain fatty acid | LC/MS neg |
| **heptadecanoate** | Lipid | Long chain fatty acid | LC/MS neg |
| **stearate180** | Lipid | Long chain fatty acid | LC/MS neg |
| **oleate181n9** | Lipid | Long chain fatty acid | GC/MS |
| **Nonadecenoate** | Lipid | Long chain fatty acid | LC/MS neg |
| **Eicosenoate** | Lipid | Long chain fatty acid | LC/MS neg |
| **Linoleate** | Lipid | Long chain fatty acid | LC/MS neg |
| **Linolenate** | Lipid | Long chain fatty acid | LC/MS neg |
| **dihomo-linoleate** | Lipid | Long chain fatty acid | LC/MS neg |
| **Adipate** | Lipid | Fatty acid, dicarboxylate | GC/MS |
| **Azelate** | Lipid | Fatty acid, dicarboxylate | LC/MS neg |
| **sebacate** | Lipid | Fatty acid, dicarboxylate | LC/MS neg |
| **dodecanedioate** | Lipid | Fatty acid, dicarboxylate | LC/MS neg |
| **octanoylcarnitine** | Lipid | Carnitine metabolism | LC/MS pos |
| **decanoylcarnitine** | Lipid | Carnitine metabolism | LC/MS pos |
| **cis4decenoylcarnitine** | Lipid | Carnitine metabolism | LC/MS pos |
| **laurylcarnitine** | Lipid | Carnitine metabolism | LC/MS pos |
| **hydroxypalmitate2** | Lipid | Fatty acid, monohydroxy | LC/MS neg |
| **hydroxystearate** | Lipid | Fatty acid, monohydroxy | LC/MS neg |
| **9-HODE** | Lipid | Fatty acid, monohydroxy | LC/MS neg |
| **Choline** | Lipid | Glycerolipid metabolism | LC/MS pos |
| **Arachidonoylglycerophosphocholine** | Lipid | Lysolipid | LC/MS neg |
| **docosapentaenoyl glycerophosphocholine** | Lipid | Lysolipid | LC/MS pos |
| **docosahexaenoylglycerophosphocholine** | Lipid | Lysolipid | LC/MS pos |
| **hydroxypregnenolonedisulfate** | Lipid | Sterol/Steroid | LC/MS neg |
| **pregnsteroidmonosulfate** | Lipid | Sterol/Steroid | LC/MS neg |
| **androsten3beta17betadioldisulfate2** | Lipid | Sterol/Steroid | LC/MS neg |
| **erythritol** | Xenobiotics | Sugar, sugar substitute, starch | GC/MS |
| **Glycine** | Amino acid | Glycine, serine and threonine metabolism | GC/MS |
| **Betaine** | Amino acid | Glycine, serine and threonine metabolism | LC/MS pos |
| **Serine** | Amino acid | Glycine, serine and threonine metabolism | GC/MS |
| **threonine** | Amino acid | Glycine, serine and threonine metabolism | LC/MS pos |
| **Alanine** | Amino acid | Alanine and aspartate metabolism | GC/MS |
| **N-acetylalanine** | Amino acid | Alanine and aspartate metabolism | LC/MS neg |
| **glutamate** | Amino acid | Glutamate metabolism | LC/MS neg |
| **pyroglutamine** | Amino acid | Glutamate metabolism | LC/MS pos |
| **Lysine** | Amino acid | Lysine metabolism | LC/MS pos |
| **glutaryl carnitine** | Amino acid | Lysine metabolism | LC/MS pos |
| **phenylalanine** | Amino acid | Phenylalanine & tyrosine metabolism | LC/MS pos |
| **phenylacetylglutamine** | Amino acid | Phenylalanine & tyrosine metabolism | LC/MS pos |
| **tyrosine** | Amino acid | Phenylalanine & tyrosine metabolism | LC/MS pos |
| **hydroxyphenyllactate** | Amino acid | Phenylalanine & tyrosine metabolism | GC/MS |
| **tryptophan** | Amino acid | Tryptophan metabolism | LC/MS pos |
| **kynurenine** | Amino acid | Tryptophan metabolism | LC/MS pos |
| **isovalerate** | Lipid | Fatty acid metabolism | LC/MS neg |
| **isovalerylcarnitine** | Amino acid | Valine, leucine and isoleucine metabolism | LC/MS pos |
| **isobutyrylcarnitine** | Amino acid | Valine, leucine and isoleucine metabolism | LC/MS pos |
| **methioninesulfoxide** | Amino acid | Cysteine, methionine, SAM, taurine metabolism | LC/MS pos |
| **aminobutyrate** | Amino acid | Butanoate metabolism | LC/MS pos |
| **hydroxybutyrate2** | Amino acid | Cysteine, methionine, SAM, taurine metabolism | GC/MS |
| **Urea** | Amino acid | Urea cycle; arginine-, proline-, metabolism | GC/MS |
| **Proline** | Amino acid | Urea cycle; arginine-, proline-, metabolism | LC/MS pos |
| **trans-4-hydroxyproline** | Amino acid | Urea cycle; arginine-, proline-, metabolism | LC/MS pos |
| **guanidinobutanoate** | Amino acid | Guanidino and acetamido metabolism | LC/MS pos |
| **oxoproline** | Amino acid | Glutathione metabolism | LC/MS neg |
| **gammaglutamyltyrosine** | Peptide | gamma-glutamyl | LC/MS pos |
| **glycylvaline** | Peptide | Dipeptide | LC/MS pos |
| **Lactate** | Carbohydrate | Glycolysis, gluconeogenesis, pyruvate metabolism | GC/MS |
| **mannose** | Carbohydrate | Fructose, mannose, galactose, starch, and sucrose metabolism | GC/MS |
| **Citrate** | Energy | Krebs cycle | GC/MS |
| **succinate** | Energy | Krebs cycle | LC/MS neg |
| **phosphate** | Energy | Oxidative phosphorylation | GC/MS |
| **heptanoate** | Lipid | Medium chain fatty acid | LC/MS neg |
| **Laurate** | Lipid | Medium chain fatty acid | LC/MS neg |
| **nonadecanoate** | Lipid | Long chain fatty acid | LC/MS neg |
| **eicosapentaenoate** | Lipid | Essential fatty acid | LC/MS neg |
| **docosahexaenoate** | Lipid | Essential fatty acid | LC/MS neg |
| **dihomolinolenate** | | | |
| **arachidonate** | Lipid | Long chain fatty acid | LC/MS neg |
| **adrenate** | Lipid | Long chain fatty acid | LC/MS neg |
| **carboxy4methyl5propyl2furanpropanoate** | Lipid | Fatty acid, dicarboxylate | LC/MS neg |
| **propionylcarnitine** | Lipid | Fatty acid metabolism (also BCAA metabolism) | LC/MS pos |
| **deoxycarnitine** | Lipid | Carnitine metabolism | LC/MS pos |
| **carnitine** | Lipid | Carnitine metabolism | LC/MS pos |
| **HETE** | Lipid | Fatty acid, monohydroxy | LC/MS neg |
| **myoinositol** | Lipid | Lysolipid | |
| **glycerophosphorylcholine** | Lipid | Glycerolipid metabolism | LC/MS pos |
| **Palmitoleoylglycerophosphocholine** | Lipid | Lysolipid | LC/MS pos |
| **oleoylglycerophosphocholine** | Lipid | Lysolipid | LC/MS pos |
| **Palmitoylplasmenylethanolamine** | Lipid | Lysolipid | LC/MS pos |
| **Palmitoylglycerophosphoinositol** | Lipid | Lysolipid | LC/MS neg |
| **Arachidonoylglycerophosphoinositol** | Lipid | Lysolipid | LC/MS neg |
| **glycerol** | Lipid | Glycerolipid metabolism | GC/MS |
| **glycerol3phosphate** | Lipid | Glycerolipid metabolism | GC/MS |
| **cholesterol** | Lipid | Sterol/Steroid | GC/MS |
| **pregnendioldisulfate** | Lipid | Sterol/Steroid | LC/MS neg |
| **Cortisol** | Lipid | Sterol/Steroid | LC/MS pos |
| **androsten3beta17betadioldisulfate1** | Lipid | Sterol/Steroid | LC/MS neg |
| **glycocholenate sulfate** | Lipid | Bile acid metabolism | LC/MS neg |
| **xanthine** | Nucleotide | Purine metabolism, (hypo)xanthine/inosine containing | LC/MS pos |
| **Urate** | Nucleotide | Purine metabolism, urate metabolism | LC/MS neg |
| **Uridine** | Nucleotide | Pyrimidine metabolism, uracil containing | LC/MS neg |
| **pseudouridine** | Nucleotide | Pyrimidine metabolism, uracil containing | LC/MS pos |
| **catecholsulfate** | Xenobiotics | Benzoate metabolism | LC/MS neg |
| **hippurate** | Xenobiotics | Benzoate metabolism | LC/MS neg |